\documentclass{aa}  
\usepackage[colorlinks=true,     linkcolor=blue, citecolor=blue, filecolor=blue, urlcolor=blue]{hyperref}
\usepackage{graphicx}
\usepackage{float}
\usepackage{txfonts}
\newcommand{\mgki}{$\rm MgI2.10$}
\newcommand{\mgkii}{$\rm MgI2.28$}
\newcommand{\feki}{$\rm FeI2.23$}
\newcommand{\fekii}{$\rm FeI2.24$}
\newcommand{\naiv}{$\rm NaI2.21$}
\newcommand{\cak}{$\rm CaI2.26$}

\newcommand{\covii}{$\rm CO2.30$}
\newcommand{\coviii}{$\rm CO2.32$}

\newcommand{\zh}{$\rm [Z/H]$}

\newcommand{\cfe}{$\rm [C/Fe]$}
\newcommand{\mgfe}{$\rm [Mg/Fe]$}
\newcommand{\ofe}{$\rm [O/Fe]$}

\newcommand{\nafe}{$\rm [Na/Fe]$}
\newcommand{\nai}{$\rm NaD$}
\newcommand{\naii}{$\rm NaI8190$}
\newcommand{\naiii}{$\rm NaI1.14$}
\newcommand{\nfe}{$\rm [N/Fe]$}
\newcommand{\cafe}{$\rm [Ca/Fe]$}
\newcommand{\tife}{$\rm [Ti/Fe]$}
\newcommand{\sife}{$\rm [Si/Fe]$}

\newcommand{\afe}{$\rm [\alpha/Fe]$}
\newcommand{\kms}{\,km\,s$^{-1}$}
\newcommand{\VROT}{$\rm V_{r}$}

\newcommand{\gammab}{$\rm \Gamma_b$}

\newcommand{\re}{$\rm R_e$}
\begin{document}
\nolinenumbers
\titlerunning{Radial gradients of K-band features in M87}
\authorrunning{F. La Barbera et al.}

   \title{Puzzling radial gradients of K-band absorption features in the giant elliptical galaxy M87}


   \author{F. La Barbera~\inst{1}, 
   A. Vazdekis~\inst{2, 3}, A. Pasquali~\inst{4}, J. Heidt~\inst{5}, E. Eftekhari~\inst{2, 3},  M.~A. Beasley~\inst{2, 3}, A. Gargiulo~\inst{6}, S. Bisogni~\inst{6}, C. Spiniello~\inst{7}, L.~P. Cassarà~\inst{6}, M. Sarzi~\inst{8} 
          }

   \institute{INAF-Osservatorio Astronomico di Capodimonte, sal. Moiariello
16, Napoli, 80131, Italy\\
              \email{francesco.labarbera@inaf.it}
         \and
             Instituto de Astrof\'\i sica de Canarias, Calle V\'\i a L\'actea s/n, E-38205
         \and
         Departamento de Astrof\'\i sica, Universidad de La Laguna (ULL), E-38206  La Laguna, Tenerife, Spain
         \and
         Astronomisches Rechen-Institut, Zentrum f\"ur Astronomie, Universit\"at Heidelberg, M\"onchhofstr. 12-14, D-69120 Heidelberg, Germany
         \and
         Landessternwarte, Zentrum f\"ur Astronomie der Universit\"at Heidelberg, K\"onigstuhl 12, 69117 Heidelberg, Germany
         \and
         INAF – Istituto di Astrofisica Spaziale e Fisica Cosmica Milano, Via A. Corti 12, 20133 Milano, Italy
         \and
         Sub-Dep. of Astrophysics, Dep. of Physics, University of Oxford, Denys Wilkinson Building, Keble Road, Oxford OX1 3RH, United Kingdom  
         \and 
         Armagh Observatory and Planetarium, College Hill, Armagh BT61 9DG, UK}

   \date{Received ; accepted}

 
   \abstract{  We  present  new  K-band  spectroscopy  for  the  giant
     elliptical galaxy M87  in the Virgo cluster, taken  with the LUCI
     spectrograph at the Large Binocular Telescope (LBT). The new data
     are used  to study  line-strengths of K-band  absorption features
     from different chemical  species, namely Fe, Mg, Ca,  Na, and CO,
     as a function of galactocentric distance, out to $\sim 40$'' from
     the  center (about  half  of the  galaxy  effective radius).  The
     radial trends of  spectral indices are compared to  those for the
     bulge of  M31, observed with  the same instrument. For  M87, most
     K-band indices  exhibit flat radial profiles,  with the exception
     of  NaI2.21,  that decreases  outwards,  with  a negative  radial
     gradient. Significant  offsets are found between  indices for M87
     and  those  for  the  bulge  of M31,  the  latter  having  weaker
     line-strengths for almost all features,  but Fe and Ca, for which
     we  find  similar trends  in  both  systems.   We find  that  the
     behavior of  CO features -  most prominent  in giant stars  - is
     difficult to  explain, consistent  with previous results  for the
     central  regions  of massive  galaxies.   In  particular, the  CO
     indices  are  stronger  in  M87  than M31,  and  do  not  exhibit
     signiﬁcant radial gradients in M87,  despite its IMF being bottom
     heavier than  M31 especially in its  central region.  Predictions
     of state-of-the-art  stellar population models, based  on results
     from the  optical spectral range, are  able to match only  the Na
     and Ca indices of M87, while  a significant mismatch is found for
     all  other indices.   This  shows  that state-of-the-art  stellar
     population models  should be  improved significantly in  order to
     provide reliable constraints on the stellar population content of
     galaxies in the NIR spectral range.  }


   \maketitle
%
\nolinenumbers
\section{Introduction}
\label{sec:intro}

Studying  absorption features  in  the spectra  of unresolved  stellar
populations  (SPs),  over  a  large wavelength  baseline,  provides  a
powerful  tool to  break degeneracies  among different  SP parameters,
such  as age,  metallicity,  {initial  mass  function (IMF)},  and
chemical  abundance  ratios.  In particular,  the  { near-infrared
  (NIR)} spectral range offers us the opportunity to study the role of
cool  stars, being  them  giants or  dwarfs,  in a  SP.   In order  to
interpret  the information  encoded in  integrated light,  one has  to
compare the  observed spectra  to predictions  of synthetic  SP models
(e.g.~\citealt{BC:03,  Vazdekis:2016,   CvD18,  Maraston:2020},  among
others).  For early-type  galaxies  (ETGs), dominated  by old  stellar
populations, this technique  has been able to  provide precise results
in the  optical, allowing us  to measure few-percent level  effects in
their  spectra, such  as variations  of the  low-mass end  of the  IMF
(\citealt{CvD12b,  LB:13,  Spiniello:2014}), of  individual  elemental
abundances  (\citealt{Serven:2005, TMJ11,  CvD12a, Vazdekis:15}),  and
how these parameters  change as a function  of galactocentric distance
in   galaxies,    out   to   the   outer,    low   surface-brightness,
regions~\citep{Greene:2015}.  Most  studies have found  that abundance
gradients  are relatively  shallow  in ETGs,  compared to  metallicity
gradients, and  that the  stellar IMF exhibits  an excess  of low-mass
stars  (i.e.   a bottom-heavy  distribution)  in  the central  regions
(e.g.~\citealt{NMN:15a,     vanDokkum:2017,     LB:19,     Sarzi:2018,
  Barbosa:2021}).

Results in the NIR are far less clear. For instance,  \citet{Smithetal:2015} analyzed the NaI1.14 feature of ETGs in the J band, concluding that this feature is underpredicted by state-of-the-art SP models, and should not be used to obtain reliable IMF constraints. A similar result has been obtained by~\citet{Benny:2017}, analyzing the NaI2.21 absorption in the K band. However, ~\citet[hereafter LB17]{LB:17} found that SP models varying simultaneously IMF slope and Na abundance are able to match  all the four prominent Na features in the spectra of massive ETGs ({ namely \nai, \naii , \naiii, and \naiv , at $\lambda \sim $0.59, 0.82, 1.14, and 2.21$~\mu$m, respectively}), across the optical and the NIR spectral range. The situation is even more debated for what concerns CO absorption features, that are prominent in the H and K bands. 
CO lines dominate the atmospheres of giant, rather than dwarf, stars. \citet{Frogel:1978, Frogel:1980} found that in ETGs, the first CO overtone in the K band, at $\sim$2.3~$\mu$m (hereafter CO2.30), was significantly stronger than expected for a { dwarf-dominated} population, { suggesting instead the presence of} a population of low-temperature luminous giants, such as (asymptotic) giant-branch (AGB) stars { (including their temporary pulsating, TP, phase), which are mostly prominent in "intermediate-age" stellar populations ($\sim$1~Gyr; ~\citealt{Conroy:2013, Maraston:2005}). For this reason, several studies have used CO line-strengths to infer the presence of intermediate-age populations in ETGs (e.g.~\citealt{MJ:1996, MJ:2000, MarmolQueralto:2009, Zibetti:2013}). A different way to scrutinize the effect of intermediate-age populations is to look at relic galaxies, i.e. massive compact ETGs that have believed to host "pristine" stellar populations, having missed recent accretion/star-formation events~\citep{Trujillo:2009, Anneta:2017, Spiniello:2021, Spiniello:2024}. As in these systems one does not expect a significant contribution from young and intermediate-age stars, one would expect weaker CO absorptions. However, 
~\citet{elham:2022b} studied NIR CO absorptions in the massive galaxy NGC\,1277 - the prototype of massive relics in the nearby Universe} - finding that CO lines in this galaxy are as strong
as for other massive ETGs. { Moreover, } ~\citet{LB:24} found that for the bulge of M31, where intermediate-age populations have been detected { \citep{Dong:2018}}, CO absorption is significantly weaker than for massive ETGs.   
{
Therefore, the  origin of the strong CO in the center  of massive galaxies remains poorly understood. More in general, it is not clear why state-of-the-art stellar population models struggle to match some NIR spectral features (see~\citealt{elham:2022a}), while they provide an excellent matching in the optical spectral range (e.g.~\citealt{LB:13}). This might be due to the presence of (cool) stars that are prominent in the NIR and do not contribute significantly to the optical, and/or to effect of stellar population properties (e.g. abundance ratios; see below) whose  effect might be difficult to model at optical wavelengths, hence motivating for detailed NIR SP studies. 
}

One major problem of SP studies at wavelengths longer than $\sim 1$~$\mu$m is to rely on accurate SP models, based on observed ("empirical") stellar spectra. The advent of new stellar libraries in the NIR, such as the (extended-) Infrared Telescope Facility~(IRTF; \citealt{IRTFI, IRTFII, Villaume:2017}) and X-Shooter spectral  library (XSL; \citealt{Verro:2022a}), has provided a significant leap forward in this field, although some libraries are affected by a dearth of high-temperature stars, hampering the computation of models at young ages ($\lesssim 1$~Gyr; see~\citealt{RV:16}).  Significant progress has also been done in the framework of theoretical stellar libraries~(see, e.g., \citealt{Allard:2012, Husser:2013}, and references therein), though the latter are hampered by large uncertainties on model atmospheres' calculations, especially for low-temperature stars (see, e.g., \citealt{Lancon:2021}). 
From an observational point of view, in order to test predictions of state-of-the-art SP models, one needs high-quality spectroscopy of galaxies in the NIR. Even for nearby systems, this is not an easy task, as ground-based observations are strongly affected by telluric absorption from the Earth's atmosphere and sky emission { lines. }

{ NIR spectroscopy exists only for sparse samples of galaxies, mostly targeting only the galaxies' central regions for both ETGs~\citep{Benny:2017, elham:2021a, elham:2022a, Gasparri:2021} and disk galaxies (see, e.g., ~\citealt{Davidge:2020})}.
Studying radial gradients of NIR absorption features is still a new, unexplored, research field, with only a few works being published so far, based on stacked spectra~\citep{Alton:2017, Alton:2018}. On the other hand, this kind of analysis provides access to a wealth of information. Several IMF-sensitive features have been identified so far in the J, H, and K bands~\citep{CvD18, elham:2021a}, allowing us, in principle, to constrain radial gradients in the low-mass end of the IMF and its functional form. The effect of age gradients is basically negligible in the NIR~\citep{elham:2021a}, compared to the optical, but for the possible contribution of cool evolved (AGB) stars from intermediate-age ($\lesssim 1.5$~Gyr) SPs. Last, but not least, variations of elemental abundances are expected to affect almost in an { orthogonal way optical and NIR features (e.g.~\citealt{CvD12a})}, allowing abundance gradients to be constrained in detail. However, in order to exploit this information, one needs to rely on SP models that are accurate enough to match NIR spectra with a percent level accuracy. To this effect, one has to  benchmark the models against high-quality galaxies' spectra. 

Stepping in this direction, we have undertaken a detailed study of NIR absorption features in nearby galaxies, using the LUCI spectrograph onboard of the Large Binocular Telescope (LBT). In~\citet[hereafter LB24]{LB:24}, we have presented new NIR spectroscopy with the LBT for the bulge of M31. Due to its promixity,  the Andromeda galaxy allows us to obtain high S/N spectroscopy in the NIR spectral range, at low observational cost,  and the stellar population analysis can piggyback on a wealth of independent constraints from spatially resolved stellar population studies, optical spectroscopy, and dynamical studies. In the present work, we make a leap forward, presenting new, high-quality, LBT spectroscopy in the K band, for the giant elliptical galaxy NGC\,4486 (M87), in the Virgo cluster. We study radial gradients of K-band absorption features, from different chemical species, namely Fe, Mg, Ca, Na, and CO, out to a galactocentric distance of $\sim 0.5$~\re , comparing the radial trends to those for the bulge of M31.  M87 is one of the brightest massive ETGs in the nearby Universe, making it the ideal target for the present investigation. A detailed analysis of the stellar population content of this galaxy, and in particular its stellar IMF, has been performed in~\citet[hereafter S18]{Sarzi:2018}, based on MUSE (optical) data, out to a galactocentric distance of $\sim 0.5$~\re , i.e. approximately the same radial range covered by our new LBT data. 
{ S18 found that M87 has old ages ($\sim 14$~Gyr) at all radii, with a metallicity gradient decreasing from supersolar values ($\sim 0.2$~dex) in the innermost region, to slightly subsolar (about $-0.1$~dex) at $\sim 0.5$~\re . Over the same radial range, the abundance of alpha elements ($\rm [\alpha/Fe]$) decreases from $\sim 0.43$ to $\sim 0.3$~dex, while $\rm [Na/Fe]$ shows a more significant gradient (from $\sim 0.7$ to $\sim 0.45$~dex).  }
Throughout the present work, we make extensive use of results from S18 in order to interpret the K-band data of M87. We adopt an effective radius of $\rm R_e = 81.2$'' for M87,  as in~\citet{Cap:11}, and assume a distance of 16.5~Mpc to this galaxy~\citep{Jordan:2007}. All wavelengths are quoted in the air system.

\begin{figure*}
\begin{center}
 \leavevmode
 \includegraphics[width=18.5cm]{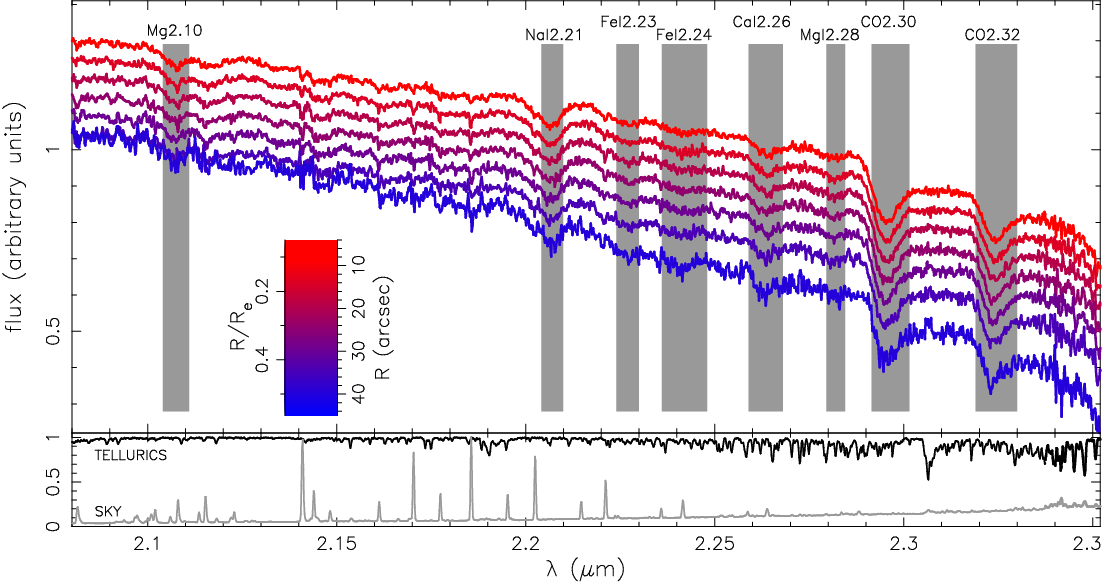}
\end{center}
 \caption{
{ Top panel:} K-band radially binned spectra obtained with LUCI@LBT for M87. Each spectrum has a S/N ratio larger than 90 (per \AA ; see the text). An arbitrary vertical shift has been applied among different spectra. The main absorption features (see Sect.~\ref{sec:indices}) are marked with gray shaded regions. Different colors correspond to different galactocentric distances, with red marking the center and blue the outermost radial bins (see the inset colored bar). 
{ All the spectra in the plot are shown at their nominal velocity dispersion, $\sigma$. Bottom panel: examples of telluric absorption (black) and sky emission (gray) spectra in the restframe of M87 (assuming an heliocentric velocity of 1284~\kms\ as in~\citealt{Cap:11}). The sky emission spectrum is normalized to a maximum value of one for displaying purposes. }
 }
   \label{fig:M87spec}
\end{figure*}

\begin{figure}[t]
\begin{center}
 \leavevmode
 \includegraphics[width=8.5cm]{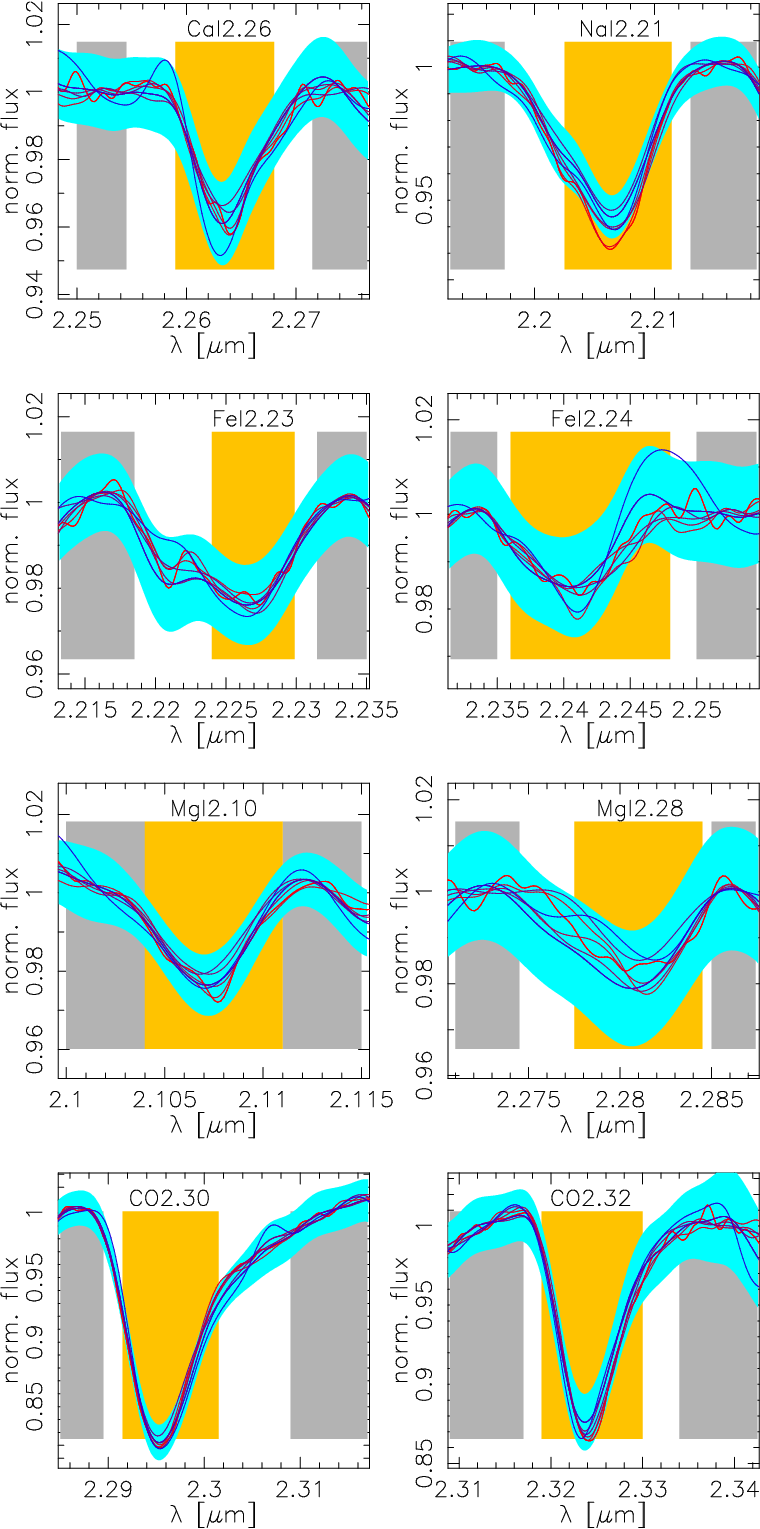}
\end{center}
 \caption{
K-band spectra of M87 in the regions where spectral indices are measured. From left to right, and top to bottom, the plots correspond to \cak\ and \naiv\ (first row), \feki\ and \fekii\ (second row), \mgki\ and \mgkii\ (third row), \covii\ and \coviii\ (bottom row), respectively.
Spectra with different colors correspond to different galactocentric distances, with the same color coding as in Fig.~\ref{fig:M87spec}. Grey regions mark the blue and red pseudocontinua of the indices, while the central passbands are marked in orange. All spectra have { been smoothed to the same $\sigma$ of 320~\kms , and } normalized by linear fit of the pseudocontinua. { Light-blue} regions are the 1~sigma uncertainties for the outermost spectrum.
 }
   \label{fig:spec_indices}
\end{figure}

\section{Data}
\label{sec:data}

\subsection{New K-band spectroscopy of M87}
\label{sec:obs}

\subsubsection{Observations}
We have obtained new long-slit spectroscopy in the K band along the major axis of M87 { (PA=-20$^{\circ}$)}~\footnote{This avoids the direction in which the jet is pointing, at PA=-69$^{\circ}$ (see e.g. S18).}, using the two NIR spectrographs, LUCI1 and LUCI2, at the Large Binocular Telescope (LBT).
The  spectra were acquired between March and July 2023, under Italian LBT open time  (proposal IT-2022B-011; PI: F. La Barbera) and German Ratszeit GTO. Observations consisted of six sequences, each with 20 individual exposures. { For each exposure, we adopted DIT=90~s and NDIT=1. 
 To perform sky subtraction, one dedicated off-target exposure was taken every two on-source exposures, resulting in a total sky exposure time half of that on-source. We adopted the same strategy and setup (with DIT =90 s) for both the target and the sky.}
Four { of the six } sequences were observed with both LUCIs simultaneously, in binocular mode, while the remaining two sequences were obtained with LUCI1 only. Small offsets (10''-wide) along the slit  were applied among different exposures, in order to allow for an optimal removal of cosmic rays and detector artifacts.
{ Remaining artifacts were efficiently removed by a clipping procedure applied during the combination of all available frames (see below).}
 After discarding low-quality frames (because of bad weather conditions, and/or strong instrumental artifacts), we remained with a total of 98 exposures, totalling an on-source exposure time of 2.45~hr.

{ The telluric standard star HD111744, of spectral type A0V, was observed before and after the science exposures, with the same setup as M87. The telluric correction was derived as follows. First, we removed the Brackett gamma line from the telluric star spectrum, by normalizing it with a multi-gaussian best-fit model of the line. Then, we derived a telluric absorption model by normalizing the cleaned spectrum of HD111744 by its continuum~\footnote{The continuum was modeled with a cubic spline, by considering only pixels not affected by telluric lines.}. The telluric correction was then performed by normalizing the science frame with a rescaled version of the telluric absorption model. } To perform relative flux calibration, we also observed the star HD106038 (spectral type F6V), from the IRTF spectral library~\citep{IRTFI}, right after the observations of M87.

Observations were carried out with the 0.5''-wide { slit, having a slit length of $\sim 210$'', hence covering} a spatial region of about $\pm$~100'' around the photometric center of M87, with a spatial scale of 0.25''/pixel. The typical seeing was $\sim$1'' (FWHM). We adopted the G210 grating and the K-band filter. Combining all the LUCI1 and LUCI2 available frames, this setup resulted  into a wavelength range of 2.02--2.36~$\mu$m, and a uniform instrumental resolution { of $\sim 35$~\kms}. This resolution, significantly higher than the galaxy velocity dispersion ($\gtrsim 250$~\kms , see App.~\ref{app:kin}), allowed us to perform an accurate correction of sky emission and telluric lines.

\subsubsection{Data reduction}
\label{sec:red}
The data were reduced using dedicated FORTRAN and IRAF routines developed by the authors, as detailed in LB24. In short, all frames were dark corrected, trimmed, and flat-fielded using dome-flats. Wavelength calibration was performed by using sky lines to optimize the wavelength solution from Ar+Ne+Xe arc-lamp frames to each individual science frame. The accuracy of the wavelength calibration is $\sim 5$~\kms\ (rms).
Relative flux calibration was performed through the observations of an IRTF star, adopted as spectrophotometric standard star.
Sky subtraction was performed by correcting each M87 frame with the nearest available sky exposure, by using the software {\sc SKYCORR}~\citep{Noll:2014}. The telluric standard stars were used to correct each exposure for telluric lines. 
The sky-corrected frames were rectified and combined. Variance maps were combined in the same way as the science data. 

In order to test the reliability of sky and telluric correction, we carefully inspected the extracted spectra of M87 at different radii (see below), finding no clear correlation between fluctuations in the data, and the position of sky/telluric lines.  We also compared line-strengths of spectral indices used in the present work at the same galactocentric distance but from opposite sides of the slit, finding consistent values within the error bars. Finally, we combined only the best-quality data of M87, from different observing periods (see above), still finding an excellent agreement { (at 1~sigma level)} in the estimated line-strengths. 

\subsubsection{Radially binned spectra}
\label{subsec:binning}
We extracted radially binned spectra along both sides of the slit. The central bin, around the photometric center of the galaxy,  was set to be 1.5$''$-wide, 
while for the other bins we adopted a minimum width of $0.75''$ ($\sim 3$ pixels), increasing it adaptively outwards, to ensure a minimum average signal-to-noise ratio, $\rm S/N_{min}$ (per \AA). In order to derive the line-of-sight kinematics, i.e. { radial} velocity \VROT , and velocity dispersion, $\sigma$, we adopted $\rm S/N_{min}=40$. 
As detailed in App.~\ref{app:kin}, the kinematics of M87 in the K band is fully consistent with that derived in the optical from previous studies. The galaxy shows no rotation, with $\sigma$ increasing from $\sim 260$~\kms\ at galactocentric distances $\gtrsim 20''$, up to $\sim 360$~\kms\ in the center.

To perform the stellar population analysis (see below),
we folded up spectra from both sides of the slit, and constructed a second set of radially binned spectra, by setting $\rm S/N_{min}=90$. This S/N ratio allows us to probe a large spatial region (approximately $\rm \pm 0.5 \, R_e$) around the galaxy center. Note that in order to { safely avoid} contamination from the 
non-thermal continuum associated to either the jet or the AGN of M87, we excluded spectra at distances below $5$'', where the featureless continuum of the jet/AGN tends to dilute the spectral indices (S18). This procedure resulted into a set of 7 binned spectra, { as shown in Fig.~\ref{fig:M87spec} (top panel). In order to illustrate the possible effect of sky residuals, the Figure also shows an example of telluric absorption and sky emission spectra for the LUCI data of M87 (bottom panel). Overall, the LUCI data have excellent quality, out to the largest galactocentric distance probed in the present work ($\rm R \sim 0.5 \, R_e$, i.e. the same radial range as the MUSE data presented in S18).
} 

\begin{table*}
\centering
\small
 \caption{Definition of spectral indices used in the present work. 
 Wavelengths are quoted in the air system. All indices are in units of \AA . }
  \begin{tabular}{c|c|c|c|c}
 Index  & Blue Pseudo-continuum & Central passband & Red Pseudo-continuum & Reference \\
        & [\AA] & [\AA] & [\AA] &      \\
   (1)  &   (2) &   (3) &   (4) & (5)  \\
   \hline
\mgki    & $21000$--$21040$ & $21040$--$21110$ & $21110$--$21150$ & \citet{Ivanov:04} \\
\mgkii    & $22710$--$22745$ & $22775$--$22845$ & $22850$--$22874$ & \citet{Silva:08}, this work  \\
\feki    & $22133$--$22185$ & $22240$--$22299$ & $22315$--$22350$ &  \citet{Silva:08}, this work\\
\fekii    & $22315$--$22350$ & $22360$--$22480$ & $22500$--$22545$ & EVL21  \\
\naiv    & $22012$--$22039$ & $22041$--$22099$ & $22100$--$22156$& LB17   \\
\cak    & $22500$--$22545$ & $22590$--$22680$ & $22715$--$22765$ &   EVL21 \\
\covii    & $22850$--$22895$ & $22915$--$23015$ & $23090$--$23170$ & EVL21 \\
\coviii    & $23090$--$23170$ & $23190$--$23300$ & $23340$--$23425$ & EVL21 \\
  \end{tabular}
\label{tab:defindices}
\end{table*}

\subsection{Data for the M31 bulge and massive ETGs}
\label{subsec:otherdata}

We compare the strength of K-band absorption features for M87 with those of other spheroidal systems, i.e. (i) the bulge of M31; and (ii) the stacked spectrum of the central regions of massive  ETGs.

The bulge of M31 has been observed with LUCI@LBT with the same instrumental setup as M87 (see~\citealt{LB:21} and LB24 for details). For the present analysis, we rely on K-band radially-binned  spectra of M31, obtained by folding up spectra from both sides of the LUCI slit. The spectra have high S/N ratio (S/N$\sim$200 per \AA ; see, e.g., figure~1 of LB24).
As shown  in LB21 (see also~\citealt{Saglia:2010}), {  most of the bulge, outside a galactocentric distance of $\sim 0.1$~\re , has approximately solar metallicity and old age ($\gtrsim 12$~Gyr), similar to massive ETGs, and an alpha-enhancement, \afe, of $\sim 0.2$~dex. In the innermost region,  inside $0.1$~\re , metallicity increases steeply to supersolar values (up to 0.4~dex) and the age decreases (down to $\sim 7$~Gyr)}. { We note that although the bulge of M31 has significantly lower $\sigma$ ($\sim 150$~\kms ) than  M87 ($\sigma \gtrsim 260$~\kms , see below), the comparison between M87 and M31  allows us to discuss, in a model independent way, the effect of  varying stellar population properties on spectral indices (see Sect.~\ref{sec:discussion}). } For the present analysis, we adopt a bulge effective radius of 1~kpc~\citep{Courteau:2011}.

We also rely on K-band spectroscopy for a sample of massive ETGs observed with the X-SHOOTER spectrograph at ESO VLT (hereafter X-Shooter galaxies, XSGs; see LB17,~\citealt{LB:19}).
The sample consists of seven massive, low-redshift (z=$0.05$), ETGs selected from the SPIDER sample~\citep{SpiderI}, which are Brightest Cluster Galaxies (BCGs), as M87, and have a central velocity dispersion between 300 and 360~\kms ,  i.e. similar to that of M87 (see below).
We use a high-S/N stacked spectrum for the central regions of the XSGs (within $\sim 1.4$~kpc, i.e. $\rm R/R_e \sim 0.04$), constructed as described in ELV22a.
As shown in LB19, the XSGs exhibit old stellar populations in their center (with an age of $\sim$11~Gyr), high metallicity (\zh\ between 0.15 and 0.4~dex), and \afe\ as high as 0.4~dex,  with a significantly bottom-heavy IMF. We note that some spectral indices (see below) are not available for the XSGs. In particular, the CO2.32 index is outside the XSG wavelength range, while \fekii\ turned out to be severely contaminated by sky residuals, and was excluded from the present analysis.

{ We point out that the bulge of M31 has lower \afe\ ($\sim 0.2$~dex) than M87 and the XSGs (whose \afe\ varies between $\sim 0.3$ and $\sim 0.4$~dex; see above and Sect.~\ref{sec:intro}). This difference implies that the bulge experienced star formation over a longer timescale, about $\sim 2$~Gyr (see figure~2 of ~\citealt{deLaRosa:2011}), than M87 and the XSGs. However, a difference of a few Gyrs is completely negligible for our purposes, as both the M31 bulge and massive ETGs have very old ages (see above), and NIR spectral indices (especially in K band) are completely insensitive to age for old populations (see, e.g., ~\citealt{elham:2022a}).}

\section{Stellar population models}
\label{sec:models}
To analyze the spectra of M87, we rely on  EMILES~\citep{Vazdekis:2016} and \citet[hereafter CvD18]{CvD18} stellar population models~\footnote{{ A comparison} between predictions of EMILES models and those of~\citet{Maraston:2005} and ~\citet[hereafter XSL]{Verro:2022a, Verro:2022b} has been already presented in~ELV22a, and  therefore, it is not shown in the present paper. In particular, XSL models have been shown to significantly underpredict the strength of CO absorptions in the NIR { (more than EMILES and CvD18 models)}, hampering the comparison between data and models.}.

EMILES  simple stellar population (SSP) models cover the spectral range from $0.168$ to $5 \, \mu$m, based on different empirical stellar libraries, namely the NGSL~\citep{Gregg:2006}, MILES~\citep{MILESI}, Indo-US~\citep{Valdes04},
CaT~\citep{CATI} and IRTF~\citep{IRTFI,IRTFII} stellar libraries (see~\citealt{Vazdekis:12},~\citealt{RV:16}). 
In the present work, we use an updated version of EMILES models, where the computation of SSPs in the NIR spectral range is based on both the IRTF and IRTF-extended~\citep{Villaume:2017} stellar libraries (see Vazdekis et al., in preparation). The two libraries are merged by interpolating  the IRTF-extended spectra to the same wavelength range as the original IRTF stars~\footnote{ Spectra from the IRTF-extended library cover the spectral range from $\sim$0.7 to $\sim$2.5~$\mu$m, while those of the IRTF library extend from $\sim$0.8 to $\sim$5~$\mu$m. Therefore, at the blue end, no wavelength gap is introduced when interpolating the extended library to the range of the non-extended one. At the red end (from 2.5 to 5~$\mu$m), spectra of the extended library were extended to 5~$\mu$m with the same procedure described in~\citet{RV:16}. However, this spectral region is not used in the present work. }. Since all the IRTF-extended stars are in common with the MILES stellar library, we rely on MILES stellar parameters, namely $\rm \log g$, $\rm T_{eff}$, and $\rm [Fe/H]$, as in the EMILES models, based on the compilation of~\citet{Cenarro:2007}. In total, 380 (175 IRTF + 205 IRTF-extended) stars are used in the updated models, providing a significantly better coverage in metallicity, especially in the subsolar metallicity range, than the original IRTF library (see, e.g., figure~1 of \citealt{Villaume:2017}).
The EMILES models are computed for two sets of scaled-solar theoretical isochrones, namely
the ones of \citet{Padova00}
(Padova00) and those of \citet{Pietrinferni04} (BaSTI), the latter having lower temperatures  at the low-mass end~(see \citealt{Vazdekis:15}, and references therein).
The SSPs are computed for ages from $\sim 0.06$ to $\sim 17.8$\,Gyr ($0.03$ to $14$~Gyr), and metallicity, \zh, from $-2.2$ to $+0.22$~dex ($-1.7$ to $+0.26$~dex), for Padova00 (BaSTI)~\footnote{Note that models based on BaSTI isochrones are also computed for \zh$=0.4$ (see \citealt{Vazdekis:2016}). However, given the lower quality of these models, they are not used in the present analysis. Moreover, predictions of all models become unsafe for ages younger than $\sim 1$~Gyr in the NIR, unlike in the optical spectral range, due to the lack of hot stars in the IRTF library.} isochrones. { We note that in the present work, we rely on models based on BaSTI isochrones, though we verified that none of our conclusions would change when relying on Padova00 isochrones (see also~\citealt{LB:24}).}
As in LB24, we use models computed for a single power-law, low-mass ($\lesssim 0.5 \, M_\odot$) tapered (``bimodal'') IMF, with a logarithmic slope, $\rm \Gamma_b$. We note that in this parametrization, increasing the high-mass end slope does also increase the dwarf-to-giant ratio in the IMF (implying a more bottom-heavy distribution) through its overall normalization. The bimodal IMF has been shown to provide mass-to-light ratios consistent with dynamical constraints~\citep{Lyubenova}.  The EMILES models are computed for \gammab= \{$0.3, 0.5, 0.8, 1.0, 1.3, 1.5, 1.8, 2.0, 2.3, 2.5, 2.8, 3.0, 3.3, 3.5$\}. For $\rm \Gamma_b=1.3$, the bimodal IMF closely approximates the~\citet{Kroupa01} IMF.
An extended version of EMILES, with varying \nafe\  (hereafter NAMILES), has been presented in~LB17, 
and is used in the present work to model the response of K-band features (in particular \naiv\ ) to varying Na abundance.


CvD18 models are an updated version of those of~\citet[hereafter CvD12]{CvD12a}. The models, covering the wavelength range 0.35--2.5~$\mu$m,  are based on the MILES stellar library in the optical, as well as the extended-IRTF stellar library in the NIR, the latter providing a better coverage in metallicity compared to the original IRTF (see above). The models are based on the MIST isochrones~\citep{Choi:2016, Dotter:2016}, and cover a range of ages (1--13.5~Gyr) and metallicities (from -1.5 to 0.2~dex). The SSPs are computed for different IMFs, using a three-segment parametrization, where one can change the slopes of the IMF between 0.1--0.5~$\rm M_\odot$ and 0.5--1~$\rm M_\odot$, while at higher masses the slope is fixed to that for a Salpeter distribution. The models also include a set of theoretical SSPs, computed for a Kroupa IMF, with varying abundance ratios for different elements, such as C, N, O~\footnote{CvD18 models vary O, Ne, and S in lock-step, referring to this variation as [$\alpha$s/Fe]. In the present paper, we assume that the main effect is that of O abundance, i.e. [O/Fe]=[$\alpha$s/Fe].}, Mg, Si, Ca, Ti, Fe, K, Na, and others. 



\section{Spectral indices}
\label{sec:indices}
The K-band spectra of M87 include several absorption features from different chemical species, such as Mg, Na, Fe, Ca, and CO (see Fig.~\ref{fig:M87spec}). We measured eight Lick-like spectral indices, namely \mgki , \mgkii , \feki , \fekii , \naiv , \cak , \covii , and \coviii , according to the definitions given in Tab.~\ref{tab:defindices}. For \fekii , \cak, and the COs, we used the definitions of~\citet[hereafter EVL21]{elham:2021a}, who optimized, and fully characterized, a large set of NIR spectral features. The \naiv\ was defined as in LB17, in order to perform a more direct comparison to results from that paper. The definition of \mgki\ is from~\citet{Ivanov:04}, while for \fekii\ and \mgkii, we modified the definitions of~\citet{Silva:08}, to account for the broadening of these features due to the (high) velocity dispersion of M87. 
{ Uncertainties on spectral indices were estimated by computing standard deviations of index values among 1000 Monte-Carlo realizations, where flux values in each input spectrum were shifted according to the corresponding uncertainties (assuming Gaussian distributions).}  
All line-strengths in this work have been computed by smoothing the spectra (both { the data of M87, M31, and XSGs, and the models) to a common $\sigma$} of $320$~\kms, i.e. the highest velocity dispersion of all radial bins of M87. { Uncertainties have been computed  before smoothing the spectra to a common $\sigma$.}

In order to characterize the variation of absorption indices with galactocentric distance, we fit the radial trend of a given index, $\rm I$, with a logarithmic linear relation:
\begin{equation}
\rm 
I = I_{0.1} + \nabla_I \cdot [ \log_{10}(R/R_e) + 1 ],
\label{eq:gradients}
\end{equation}
where $\rm I_{0.1} $ provides the value of the index at $\rm 0.1 \, R_e$, while $\rm \nabla_I$ is its radial gradient. The fits are performed with a robust fitting procedure, minimizing the absolute deviation of residuals along the orthogonal direction to the best-fitting line. Uncertainties on $\rm I_{0.1} $ and $\rm \nabla_I$ are obtained with 1000 bootstrap iterations where index values are randomly shifted according to their uncertainties.

\begin{figure}
\begin{center}
 \leavevmode
 \includegraphics[width=8.8cm]{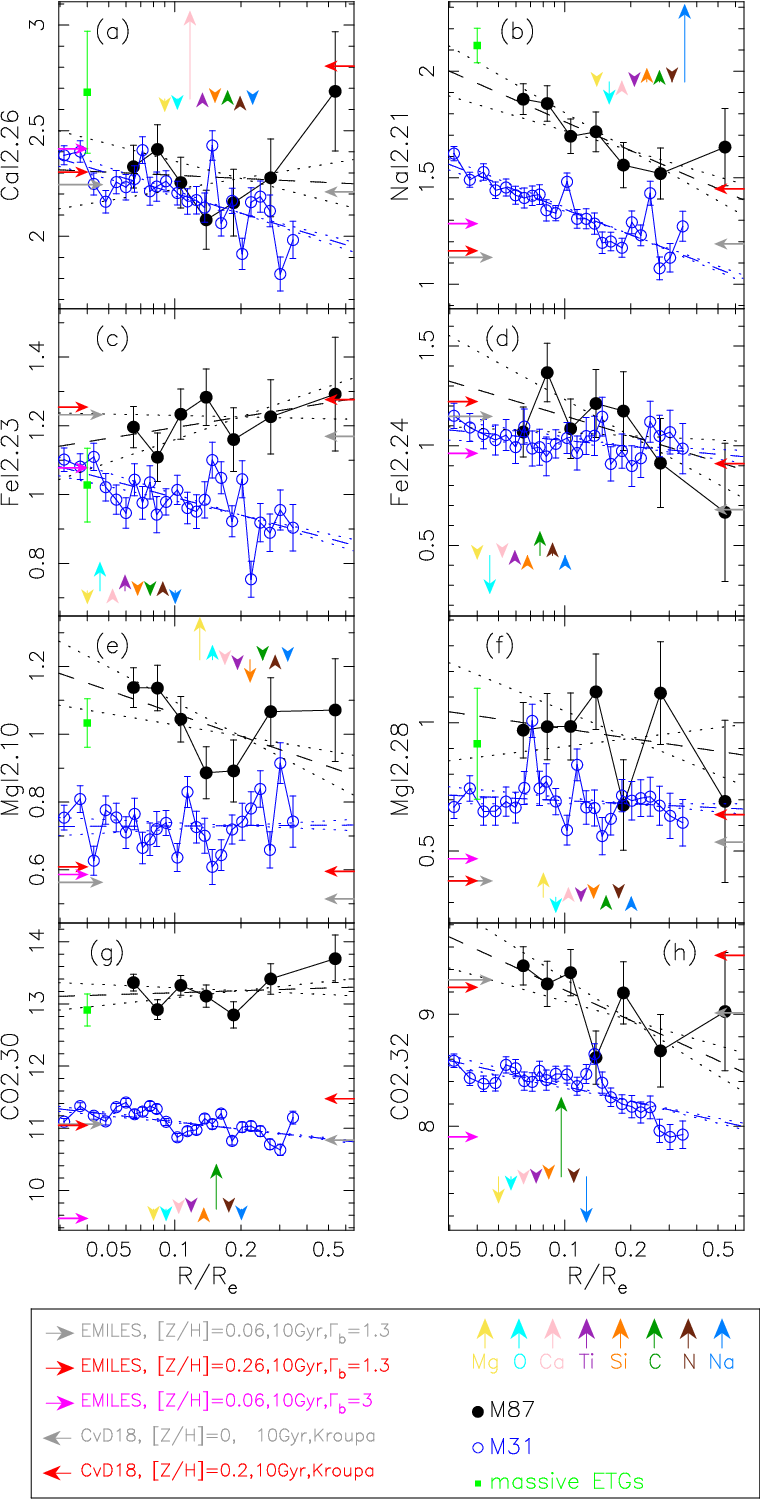}
\end{center}
 \caption{
K-band line-strength indices as a function of normalized galactocentric distance, $\rm R/R_e$. From left to right, and top to bottom, the Figure plots the same indices as in Fig.~\ref{fig:spec_indices}. { The legend in the bottom panel indicates the symbols used in the plot.} Black filled dots with (1-sigma) error bars plot the indices of M87. Data for massive ETGs and the bulge of M31 are plotted with green filled squares and blue { open} circles, respectively.
Black and blue dashed lines { represent the best-fitting radial trends for M87 and M31, respectively.
  Black and blue dotted lines show the $\pm 1 $sigma uncertainties on the slopes of the linear relations.
In each panel, predictions of EMILES-BaSTI (CvD18) models are plotted as small, rightwards (leftwards), horizontal arrows. Different colors indicate models with different metallicity and IMF slope, as indicated in the legend. }
Index changes due to variations of individual abundance ratios in CvD18 models are plotted with colored vertical arrows, showing the effect of increasing each element by +0.3~dex, with the exception of C, that is increased by 0.15~dex. 
 }
   \label{fig:indices_models}
\end{figure}

\section{Results}
\label{sec:results}

\subsection{Radial variation of K-band indices}
\label{sec:indrad}
Figure~\ref{fig:spec_indices} shows the spectra of M87 in the regions of the K-band spectral indices. The color coding of the spectra is the same as in Fig.~\ref{fig:M87spec}, with red (blue) corresponding to the innermost (outermost) radial bins.  Grey and orange filled regions in the Figure mark the pseudocontinua and central passbands of the indices (see Tab.~\ref{tab:defindices}).
For each index (panel), the spectra are normalized by the line passing through the blue and red pseudocontinua. Overall, the indices show little variation with galactocentric distance, with the only exception of \naiv , where the central spectrum (red) goes significantly deeper than the outer one, indicating that the spectral index becomes stronger towards the galaxy center. This is shown more clearly in Fig.~\ref{fig:indices_models}, plotting index values for M87 (black circles with error bars) as a function of galactocentric distance. As a comparison, we also plot the central stacked spectrum of massive ETGs observed with X-Shooter (XSGs; green rectangles with error bars; see Sect.~\ref{subsec:otherdata}), and the index values for the bulge of M31 (blue empty circles). The radial trends for M87 and M31 are fitted with logarithmic linear trends (dashed lines; see Sect.~\ref{sec:indices}), whose best-fitting coefficients are given in Tab.~\ref{tab:indcoeff}. For each panel, we also show the predictions of EMILES-BaSTI and CvD18 SSP models (see Sect.~\ref{sec:models}), marked by small horizontal arrows, as well as the effect of varying individual elemental abundances with CvD18 models (see vertical arrows with different colors in each panel, with labels in the bottom part of the Figure).
In the following, we discuss qualitatively the behavior of K-band spectral indices, postponing a more quantitative analysis to Sect.~\ref{sec:optcomp}.

\subsubsection{Mg indices}
\label{sec:Mgind}

For both M87 and M31, the \mgki\ and \mgkii\ indices behave in a similar way, with a flat radial trend (see panels e and f of Fig.~\ref{fig:indices_models}), despite the galaxy metallicity gradient in both systems. The lack of radial gradient is consistent with the fact that both EMILES and CvD18 models predict a weak dependence of  \mgki\ and \mgkii\ on metallicity.  Both indices have higher values for M87 than for M31, possibly because Mg abundance is significantly higher for M87 than for M31. Indeed, S18 measured a negative radial gradient of \mgfe\ 
for M87, with values of $\sim 0.4$~dex in the center, and $\sim 0.3$~dex at $\rm \sim 0.4 R_e$; while the bulge of M31 has \mgfe $\sim 0.2$~dex at all radii (see LB21). However, CvD18 models predict that both \mgki\ and \mgkii\ should increase only mildly with \mgfe\ (see yellow vertical arrows in panels e and f), and that \mgki\ (\mgkii ) should decrease as the abundance of \sife\ (\ofe ) increases. Therefore, based on CvD18 models, one would expect a very small difference among M31, M87, and the predictions of solar-scaled models, in contrast to the observations. We come back discussing this discrepancy in Sect.~\ref{sec:optcomp}.

\subsubsection{Na index}
\label{sec:Naind}
For M87, \naiv\ is the only K-band index showing a negative radial gradient at high significance level ($\sim$3~sigma; see Tab.~\ref{tab:indcoeff} and Fig.~\ref{fig:spec_indices}). The gradient for M31 is also negative, with a slope of $-0.40 \pm 0.03$, fully consistent with that for M87. However, the trend for M87 is significantly offset to higher values compared to M31. Note that the (high) \naiv\ for the stacked spectrum of massive ETGs (green square in the panel b) is qualitatively consistent with the extrapolation of the linear fit for M87 to smaller radii. CvD18 models predict that the index increases (decreases) with \nafe\ (\ofe ), with negligible contribution from other elements (mostly \sife , see also \citealt{Benny:2017}). The index increases with metallicity significantly more for CvD18, than for EMILES, models.
Also, \naiv\ gets stronger for models with a more bottom-heavy IMF (see the horizontal pink arrow in panel b). Therefore, the negative gradient for \naiv\ can be explained, in principle, by an increase of metallicity, \nafe , and IMF slope towards the galaxy center (see LB17,~\citealt{Benny:2017}).

\subsubsection{Fe indices}
\label{sec:Feind}
For M31, \feki\ shows a negative radial gradient, while the gradient for \fekii\ is also negative, but not statistically significant within the errors (see panels c and d of the Figure and Tab.~\ref{tab:indcoeff}). For M87, none of the Fe indices shows a significant radial gradient, though for \feki , combining the data for M87 and the massive ETGs' stack seems to suggest a positive radial gradient. Both Fe indices are expected to decrease with metallicity, with the trend being stronger for CvD18 than EMILES models, and decrease with IMF slope. Hence, the lack of a negative gradient for the Fe indices of M87 might be due to the competing effect of IMF slope and metallicity. 
CvD18 models predict that \feki\ should increase with \ofe\ (and to less extent with \sife ). This might explain the offset towards higher values of \feki\ for M87 compared to M31, as the latter system is less enhanced in alpha elements compared to M87. On the contrary, \fekii\ is expected to decrease with \ofe , and increase (to less extent) with \cfe\ and \nfe . The combined effect might explain the lack of significant difference for \fekii\ between M31 and M87.

\subsubsection{Ca index}
\label{sec:Caind}
The \cak\ index shows a negative radial gradient for the bulge of M31, while no significant gradient is detected for M87 (see Tab.~\ref{tab:indcoeff}). The result for M87 is consistent with that of~\citet{Alton:2018}, who found a logarithmic radial gradient of $-0.08 \pm 0.09$ from stacked spectra of nearby massive ETGs.
 According to CvD18 models, \cak\ increases significantly with metallicity and \cafe .  Also, based on EMILES models, the index slightly increases with IMF slope~\footnote{For these reasons, as already pointed out by~\citet{CvD12a}, combining \cak\ with the calcium triplet should provide a powerful tool to disentangle the effect of IMF and Ca abundance.}. Since Ca is expected to follow Fe for massive galaxies (see, e.g.,~\citealt{Vazdekis:1997}, \citealt{JTM:2012}), and this result does also apply to the bulge of M31~\citep{LB:25}, the negative gradient of \cak\ for M31 might result from the galaxy metallicity gradient. However, the value of \cak\ from CvD18 models with supersolar metallicity is too high with respect to that for M87 and M31. Moreover, the interpretation is hampered by the fact that for EMILES models, the sensitivity of \cak\ (as that of all K-band indices) to metallicity is almost negligible. Note also that the high \cak\ index for the stack of massive ETGs suggests that a negative radial gradient might be present also for M87, but not detected because of the large uncertainty on \cak\ for the outermost radial bin ($\rm R/R_e \sim 0.5$). Further data might help to elucidate these issues.

\subsubsection{CO indices}
\label{sec:COind}
A remarkable result from the present work is the lack of any significant radial gradient for the CO2.30 index of M87 (see panel g of Fig.~\ref{fig:indices_models}). The average value of the index is $\sim 13.2$\AA, fully consistent with that for the stack of massive ETGs, and significantly above the predictions of EMILES and CvD18 models. This result is consistent with previous studies (see~\citealt{elham:2022a} and references therein), finding that CO lines are strong in the center of massive galaxies. Remarkably, for M87, this "CO offset" is independent of galactocentric distance. For the bulge of M31, CO2.30 is significantly lower than for M87, with a very shallow negative radial gradient. According to CvD18 models, the CO2.30 line increases with \cfe , and slightly decreases with the abundance of alpha elements (mostly \mgfe\ and \ofe ). Since the abundance of \cfe\ for the bulge of M31 is $\sim 0.15$~dex, similarly to what is measured for massive galaxies (e.g.~\citealt{vanDokkum:2017, FLF:21}), a difference in C abundance seems unable, by  itself, to explain the offset between M31 and M87 (see Sect.~\ref{sec:discussion}). The same results apply to CO2.32, that shows a negative radial gradient for both M87 and the bulge of M31, with an offset to higher values for M87. 
We note that for the bulge of M31, the radial variation of CO2.32 is almost negligible within $\rm \sim 0.15 R_e$, while it steepens at larger galactocentric distances (see also LB24). 
According to CvD18 models, CO2.32 is anti-correlated with \nafe . Therefore, since M87 has a stronger Na abundance than the bulge of M31 (see~S18, LB21), the intrinsic offset in CO2.32 absorption between the two systems is even more significant than what seen in Fig.~\ref{fig:indices_models} (panel h). Finally, and perhaps more interesting, we note that both CO2.30 and CO2.32 are expected to get { considerably} weaker for a more bottom-heavy IMF, as CO absorption is prominent in the atmosphere of giant, relative to dwarf, stars~\citep{Frogel:1978}. However, M87 is offset to higher CO values, independent of galactocentric distance, despite its IMF in the center being more bottom-heavy than that of M31 (see S18, LB21).  

\begin{figure}[H]
\begin{center}
 \leavevmode
 \includegraphics[width=8.8cm]{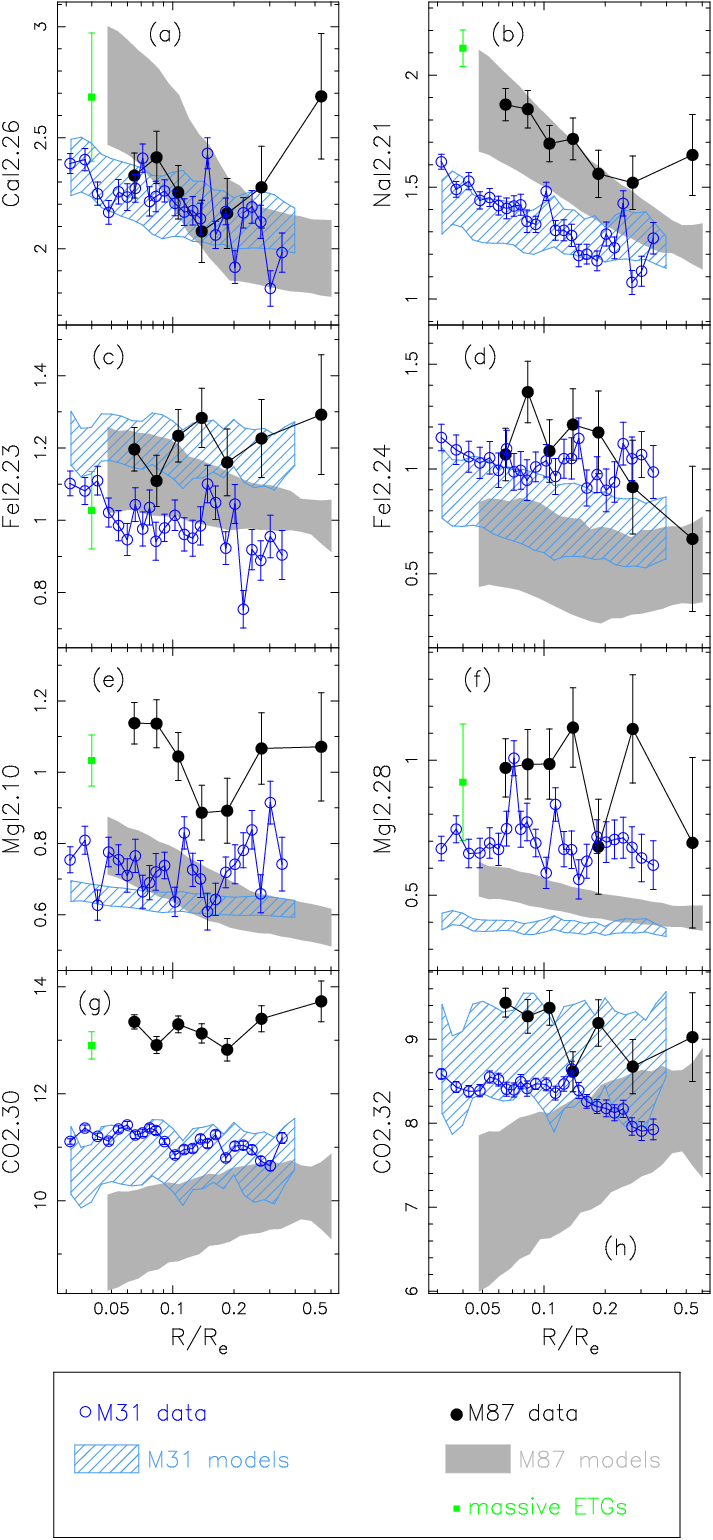}
\end{center}
 \caption{
Same as Fig.~\ref{fig:indices_models} but plotting also the range of EMILES-BaSTI SSP models' predictions based on results (age, metallicity, IMF, and abundance ratios) from the optical spectral range. 
Predictions for M31 and M87 are plotted as hatched { light-blue} and solid gray shaded regions, respectively (see the text for details). Note that, with the exception of the outermost radial bin for M87 (with larger error bars), predictions from the optical are consistent with observed line-strengths for \naiv\ and \cak\ (top panels), for both M31 and M87. For the Fe indices, there is some disagreement (at 2-3~sigma level) between data and models, while large deviations are found for Mg and CO indices (see the text).
 }
   \label{fig:comp_opt}
\end{figure}

\subsection{Comparison to predictions from the optical}
\label{sec:optcomp}

Figure~\ref{fig:indices_models} shows that state-of-the-art SSP models are not able to match (some) K-band spectral indices (e.g. CO and Mg). For this reason, we do not attempt to fit the indices with model predictions, but instead, we compare K-band line-strengths with predictions based on results from the optical spectral range. 
Figure~\ref{fig:comp_opt} shows the radial variation of K-band indices  for M87 and M31, as in Fig.~\ref{fig:indices_models}. The hatched { light-blue} (M31) and solid gray (M87) regions show a range of predictions for EMILES SSP models (with BaSTI isochrones), whose parameters (age, metallicity, IMF slope, abundance ratios) are varied according to results from the optical. The parameters are chosen as follows.  

For M87, we assume the average radial trends of age, metallicity, IMF slope (adopting a low-mass tapered IMF functional form), and \nafe\ from S18. For \mgfe, we adopt the same radial trend as \afe\ from S18~\footnote{Indeed, S18 estimated \afe\ based on Mg and Fe absorption features, mostly reflecting \mgfe\ abundance.} (see their figure~9). For the other abundance ratios, we assume average values consistent with those derived for most massive ETGs by~\citealt{CGvD:2014}. More in detail, we assume that \ofe\ is 0.1~dex above Mg (as also found for the bulge of M31; see LB24); \sife\ follows \mgfe\ (as also found in the bulge of the MW, at solar and super-solar metallicity; see~\citealt{Bensby:2017}); \nfe\ follows \mgfe ; Ti and Ca scale with Fe (i.e. \tife$ \sim 0$ and \cafe$ \sim 0$; the latter result being consistent with previous works); and a radially constant \cfe\ abundance of 0.2~dex (consistent with the best-fitting results of S18). Uncertainties on abundance ratios (but for \nafe ) are assumed to be $0.05$~dex, while for other parameters we take uncertainties from S18. In order to derive the range of predictions in Fig.~\ref{fig:comp_opt}, at each radial position, we perform several iterations where we shift each parameter by $\pm 1$~sigma (where sigma is the corresponding uncertainty), we compute the corresponding SSP predictions, and take the minimum and maximum value for each index. 
We note that although some of the above assumptions on abundance ratios are uncertain (e.g. the radial behavior of \tife; see~\citealt{Parikh:2019}), CvD18 models predict that K-band indices are only weakly dependent on most abundance ratios. Therefore, our conclusions are not affected significantly by the above assumptions.
In the case of M31, the range of predictions in Fig.~\ref{fig:comp_opt} ({ light-blue} hatched region) is derived with the same approach as for M87. Radial trends of age, metallicity, IMF slope, \mgfe, \cfe, \ofe, and \nafe\ are taken from LB21 and LB24. For the other abundance ratios, we assume \tife$ \sim $0 and \cafe$ \sim $0, \nfe $\sim $\ofe, and \sife $\sim$\mgfe\ (consistent with the analysis of the optical spectroscopy; see LB25).

Figure~\ref{fig:comp_opt} can be interpreted as follows.
\begin{description}
\item[{ Ca and Na indices} - ] For both \cak\ and \naiv , the optical solution predicts a negative radial gradient, matching reasonably well the data, except for the outermost radial bin, where both \cak\ and \naiv\ are too high compared to the models. This might be due to the large uncertainty in the outermost bin, where correction of sky lines and telluric absorption is hampered by the lower S/N ratio of unbinned spectra. Further data should allow us to address this issue.
\item[{ Mg indices} -] For both \mgki\ and \mgkii , the models are too low compared to the data, for both M31 and M87. This suggests that the response to \mgfe\ of the Mg indices might be underestimated by CvD18 models, which is consistent with the fact that the data for M87 are systematically above those for M31, the latter system having lower \mgfe\ (see Sect.~\ref{sec:data}).
\item[{ Fe indices} -] For both \feki\ and \fekii , the optical solution implies a relatively flat radial trend, consistent with the data. However, significant offsets exist between models and data. For \fekii, the model predictions tend to be lower than the observed indices for both M31 and M87. The main reason is that CvD18 models predict \fekii\ to decrease with \ofe , while the data seem to be inconsistent with this trend. 
For \feki , results are even less clear, as the models under-(over-)predict the observed indices for M87 (M31), especially for the outermost radial bins.
\item[{ CO indices} -] For the bulge of M31, the models match the observed indices, consistent with results from LB24. On the contrary, for M87, the models significantly underpredict the observed line-strengths (see also Sect.~\ref{sec:COind}), especially for the CO2.30 index, where differences are highly significant, given the error bars.   
\end{description}

\begin{table}
\centering
\small
 \caption{Coefficients of the linear fits to the logarithmic radial variation of spectral indices with galactocentric distance, $\rm R/R_e$, for M87 and for the bulge of M31 (in parentheses). For each index (Column~1), the offset, $\rm I_{0.1}$, of the relation (at $\rm R=0.1 R_e$), and its slope, $\rm \nabla_I$, are provided in Columns~2 and~3, respectively. Uncertainties are quoted at the 1-sigma level. }
  \begin{tabular}{c|c|c}
 Index  & $\rm I_{0.1}$ & $ \rm \nabla_I $ \\
   \hline
MgI2.10 & $   1.06 \pm  0.12 $ ( $  0.73 \pm  0.04$) & $  -0.22 \pm  0.13 $ ($   0.00 \pm  0.03$) \\ 
NaI2.21 & $   1.77 \pm  0.14 $ ( $  1.35 \pm  0.04$) & $  -0.45 \pm  0.15 $ ($  -0.40 \pm  0.03$) \\ 
FeI2.23 & $   1.19 \pm  0.14 $ ( $  0.99 \pm  0.04$) & $   0.10 \pm  0.14 $ ($  -0.17 \pm  0.03$) \\ 
FeI2.24 & $   1.15 \pm  0.25 $ ( $  1.03 \pm  0.06$) & $  -0.33 \pm  0.26 $ ($  -0.10 \pm  0.06$) \\ 
CaI2.26 & $   2.29 \pm  0.23 $ ( $  2.20 \pm  0.05$) & $  -0.05 \pm  0.24 $ ($  -0.31 \pm  0.05$) \\ 
MgI2.28 & $   0.98 \pm  0.24 $ ( $  0.70 \pm  0.04$) & $  -0.13 \pm  0.24 $ ($  -0.04 \pm  0.04$) \\ 
CO2.30 & $  13.18 \pm  0.29 $ ( $ 11.10 \pm  0.06$) & $   0.11 \pm  0.29 $ ($  -0.39 \pm  0.06$) \\ 
CO2.32 & $   9.22 \pm  0.41 $ ( $  8.36 \pm  0.07$) & $  -0.90 \pm  0.42 $ ($  -0.44 \pm  0.06$) \\ 
  \end{tabular}
\label{tab:indcoeff}
\end{table}

\section{Discussion}
\label{sec:discussion}

\subsection{The NaI2.21 absorption line}
\label{sec:naiv}
In the present work, we find a negative radial gradient of the \naiv\ feature, for both M87, and the bulge of M31. This result is consistent with that of~LB17, for two massive ETGs, and for stacked spectra of massive nearby galaxies from~\citet{Alton:2018}. As discussed in Sect.~\ref{sec:intro}, the explanation of Na absorption lines in the spectra of massive ETGs is a long-standing problem. While the optical NaD line is possibly affected by interstellar absorption, both in galaxy spectra~\citep{Jeong:2013} and stellar population { models~\citep{Rubin:2025}}, the NIR, gravity-sensitive,  Na lines (namely \naii , \naiii, and \naiv) are virtually unaffected by dust, and should provide a cleaner estimate of relevant stellar population parameters, such as, in particular, the low-mass end slope of the IMF and the Na abundance.  
However, ~\citet{Smithetal:2015} and~\citet{Benny:2017} found that state-of-the-art stellar population models underestimate significantly the J- and K-band Na lines (\naiii, and \naiv ). LB17 found that the discrepancy seems to be solved when using stellar population models (Na--EMILES) varying simultaneously both IMF slope and Na abundance. Figure~\ref{fig:comp_opt} (panel b) is consistent with this finding, as we are able to match the data of both M87 and the bulge of M31, at different galactocentric distances, using the stellar population properties inferred from the optical spectral range.
Therefore, the radial trend of \naiv\ is due to a combined effect of metallicity, IMF slope, and \nafe\ gradients.
We remark that if the sensitivity of \naiv\ to alpha elements (and in particular \ofe , see Fig.~\ref{fig:indices_models}) were negligible, as suggested by LB17, the agreement between models and data in panel b of Fig.~\ref{fig:comp_opt} would be even better, as the model predictions would get slightly shifted upwards (by $\sim 0.1$~\AA , for M87).  We also note that \nafe\ reaches values up to $\sim 0.7$~dex and $~0.5$~dex, in the center of M87 (see S18) and M31 (see LB21), respectively. These high overabundances do likely result because Na is primarily produced in massive stars and ejected in Type II supernovae, with metallicity dependent yields~\citep{Kobayashi:2006}, and further increased by the possible contribution of intermediate-mass AGB stars (see the discussion in LB17 and S18). 

\subsection{The unexplained trend of CO lines}
\label{sec:COpuzzle}
The strong CO absorption in the spectra of M87, and the lack of a CO radial gradient, are the most remarkable results of the present work. If the strong CO were due to low-temperature luminous giants, such as { AGB and carbon stars, whose NIR light contribution is significant in intermediate-age ($\sim 1$~Gyr) populations~\footnote{ In K band, the contribution of AGB stars to the integrated light is significant only in the age range from $\sim 0.3$ to $\sim 2$~Gyr, while at older ages the contribution of these stars to integrated light is outshined by that of other cool giant stars (mostly RGB; see figure~13 of~\citealt{Maraston:2005}). This is the main reason why a strong CO  has been mostly related to the presence of intermediate-age populations in ETGs (see Sect.~\ref{sec:discussion}).} , one would expect a more prominent effect} in the galaxy center, where a small fraction of young stars could form from recently accreted and/or left-over gas in the deeper galaxy central potential well, consistent with results from the UV spectral range~\citep{SR:2021}. This scenario is inconsistent with the K-band data for M87, where no CO radial gradient is observed, { and it is also at odds with the very old population inferred at all radii from the MUSE optical spectroscopy (S18)}. Moreover, ~\citet{elham:2022a, elham:2022b} found that models with an enhanced AGB component are not able to match the strong CO of massive ETGs, and that the massive relic galaxy NGC\,1277, where an intermediate-age population is not expected to be in place, has similar CO line-strengths as other massive galaxies.  { LB24 also } showed that for the bulge of M31, several CO indices in H and K band can be matched by SSP models without any intermediate-age component. { This would thus exclude that the CO mismatch between massive ETGs and model predictions is due to some systematics in the models (e.g. uncertainties in the estimates of stellar parameters for cool giant stars), as in this case all data (i.e. both M31 and massive ETGs) should be offset in a similar way with respect to the models.}
{ Moreover, we emphasize that since the offset is still in place in the outermost region of M87, where the IMF is similar to a Milky-Way-like distribution, the CO mismatch is not related to the presence of a bottom-heavy IMF in the center of massive ETGs. Although one could increase the CO line-strength in the models by flattening the IMF in the region around $1$M$_\odot$ (rather than steepening it) - thereby enhancing the contribution of latest evolutionary stages (RGB and AGB) in an old population -  this approach would yield only a modest increase in the CO2.30 index ($\sim 0.5$~\AA ), provided that one has to maintain the agreement between models and data in the optical spectral range.}
If the strong CO were due to \cfe\ abundance, mostly in the super-solar metallicity regime, as suggested by LB24 (see also~\citealt{elham:2022a}), one would expect no difference between M31 and M87 at a radius of $\rm \sim 0.2 \, R_e$, where both systems have similar (solar) metallicity, and no strong difference in \cfe\ is expected; in contrast to what seen in panel g of Fig.~\ref{fig:comp_opt}. The difference is even more puzzling considering that the bulge of M31 has a similar abundance pattern for several chemical elements as most massive ETGs~(LB25), suggesting a quick and fast formation process, as for most massive galaxies.
 Based on the present analysis, the origin of CO trends remains unconstrained. We speculate that massive ETGs, such as M87, might host a peculiar population of (low-temperature) stars, with strong CO, which is absent from current stellar population models. Such a population has to be a unique imprint of the formation process of these galaxies, relative to that of other systems such as the bulge of M31.  Recently,~\citet{Renzini:2023} pointed out that the unexpected abundance of high-redshift, ultraviolet-luminous galaxies, revealed by the James Webb Space Telescope (JWST), might reflect a feedback-free initial stage of star formation. One may wonder if chemical enrichment in such a formation stage, might be somehow connected to the strong CO of massive ETGs. Further work, focusing on radial gradients of massive galaxies over a wider mass range, and at larger galactocentric distances, might help to shed light on the complex origin of the CO absorption.

\subsection{Stellar population models in the NIR}
\label{sec:SPNIR}
Our analysis shows that state-of-the-art stellar population models are not able to match a number of K-band absorptions, such as CO (see above) and Mg. In the latter case, one possible explanation is that the response to varying \mgfe\ is uncertain in the models, with both \mgki\ and \mgkii\ being far more sensitive to Mg abundance than what the models predict. This might actually be the case, as K-band absorptions are mostly driven  by low-temperature (giant) stars, whose stellar atmospheres are notoriously difficult to model.
At low temperatures, differences among theoretical stellar spectra from different codes become important~\citep{Knowles:2019}, and the omission of non-LTE and 3D geometry effects in the stellar atmosphere calculations lead to larger differences with respect to data. In particular,~\citet{Lancon:2021} found that theoretical stellar spectra are a good representation of the empirical ones only down to a temperature of $\sim 4000$--$5000$~K, i.e. the regime below which stars shine the most in the NIR spectral range. For this reason, one should be extremely cautious when using K-band spectral indices, such as Mg and CO, to constrain the stellar population properties of galaxies, as this might lead to incorrect results. 

\section{Summary}
\label{sec:summary}
In the present work, we have studied K-band absorption features for the giant elliptical galaxy M87, out to a galactocentric distance of about half of the galaxy effective radius. The radial profiles are compared to those for the bulge of M31. Our results can be summarized as follows:
\begin{description}
\item[-] For M87, all K-band absorptions exhibit flat radial trends, with the exception of \naiv , showing a significantly negative radial gradient; for the bulge of M31, radial variations are small, with most indices being systematically offset to lower values compared to M87. In the center, the line-strengths of M87 are broadly consistent with those for other massive ETGs;
\item[-] The CO2.30 absorption shows a flat radial trend for both M87 and the M31 bulge. Despite having a more bottom-heavy IMF and similar metallicity, M87 shows significantly {\it stronger} CO, compared to the bulge of M31;
\item[-] Adopting stellar population parameters (namely, age, metallicity, IMF slope, and abundance ratios) derived from the optical spectral range, we show that stellar population models are able to match \naiv\ and \cak\ for both M87 and M31. However,  models underpredict significantly Mg features for both M31 and M87; while they are able to match the COs only for M31, underpredicting both CO2.30 and CO2.32 for M87.
\end{description}
We speculate that massive ETGs, such as M87, host a population of CO strong stars, which is missing in other systems, such as the bulge of M31, and is possibly connected to the unique physical conditions by which these galaxies formed at high redshift. Our study points to the importance of improving significantly the predictions of stellar population models in the NIR spectral range.

\begin{acknowledgements}

{ We thank the anonymous referee for carefully reading our manuscript and for his/her helpful comments, that significantly helped us to improve this work.}
F.L.B., A.P. and E.E. acknowledge support from the INAF minigrant 1.05.23.04.01. AV and MAB acknowledges support from grants PID2021-123313NA-I00 and PID2022-140869NB-I00 from the Spanish Ministry of Science and Innovation. 
This work has also been supported through the IAC project TRACES, which
is partially supported through the state budget and the regional budget of the Consejer{\'{i}}a de Econom{\'{i}}a, Industria, Comercio y Conocimiento of the Canary Islands Autonomous Community.
The present paper is based on LUCI-LBT spectroscopy acquired under LBTB and LBTI time. The LBT is an international collaboration among institutions in the United States, Italy and Germany. LBT Corporation Members are: The University of Arizona on behalf of the Arizona Board of Regents; Istituto Nazionale di Astrofisica, Italy; LBT Beteiligungsgesellschaft, Germany, representing the Max-Planck Society, The Leibniz Institute for Astrophysics Potsdam, and Heidelberg University; The Ohio State University, representing OSU, University of Notre Dame, University of Minnesota and University of Virginia. The data for massive ETGs have been acquired with ESO Telescopes at the Paranal Observatory under programmes ID 092.B-0378, 094.B-0747, 097.B-0229 (PI: FLB).

\end{acknowledgements}

%
%

\begin{appendix} 

  \section{Kinematics of M87 in K band}
\label{app:kin}
\begin{figure}
\begin{center}
 \leavevmode
 \includegraphics[width=8cm]{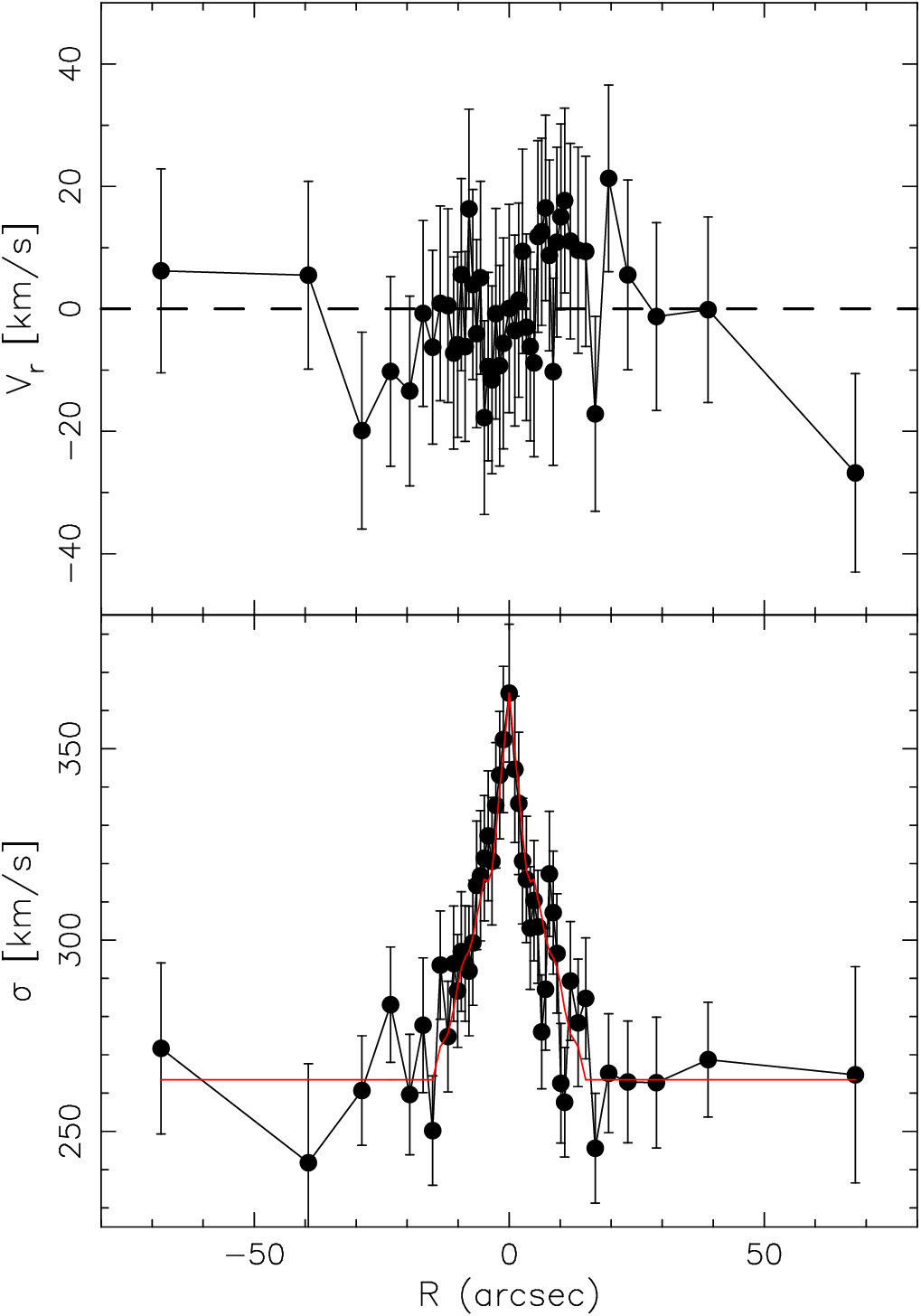}
\end{center}
 \caption{
{  Radial velocity, \VROT\ (upper panel) , and velocity dispersion, $\sigma$ (lower panel), for M87, as a function of galactocentric distance, $\rm R$ (in units of arcsec).
We have corrected \VROT\ for the system heliocentric velocity of 1284~\kms\ (as in~\citealt{Cap:11}). }
Negative and positive values of R correspond to opposite sides of the LUCI slit. The black dashed line in the upper panel marks the value of zero. Error bars correspond to 1-sigma uncertainties. In the lower panel, the red curve shows how we model the  $\sigma$ radial trend. 
 }
   \label{fig:kin_emiles}
\end{figure}

We measured the K-band kinematics of M87, { namely radial velocity, \VROT }, and velocity dispersion, $\sigma$ , using adaptively binned spectra with a minimum $\rm S/N$ ratio of 40~\AA$^{-1}$ (see Sect.~\ref{subsec:binning}), running the software {\sc pPXF}~\citep{Cap:2004, Capp:17}.
{ The pPXF fits were performed over the wavelength range from 2.16 to 2.3~$\mu$m, where most of the K-band absorption features are located (see Fig.~\ref{fig:M87spec}). We ran pPXF by including an additive polynomial of degree 5. Increasing the polynomial was found to have no significant impact on the \VROT\ and $\sigma$ estimates.}
{ Notice that $\sigma$ is obtained by adding up in quadrature the resolution of the EMILES models ($\rm \sigma_{EMILES}=60$~\kms ) to the pPXF output value, and subtracting in quadrature the LUCI instrumental resolution, $\rm \sigma_{EMILES}=35$~\kms\ (see Sect.~\ref{sec:data}).}
Fig.~\ref{fig:kin_emiles} shows the kinematic profiles obtained by feeding pPXF with EMILES 1SSP models, based on { BaSTI isochrones}~(see Sect.~\ref{sec:models}). No significant differences were found when using models with Padova isochrones. The Figure shows that M87 { has constant \VROT , implying little or negligible rotation}, while $\sigma$ steeply increases in the center, within a region of $\sim 20$''. These results are consistent with previous findings in the optical spectral range~(e.g.~\citealt{ECP:04, EKS:14}). For the present work, we have { assumed \VROT$=0$~\kms}, and approximated the $\sigma$ profile of M87 as indicated with the red curve in the lower panel of Fig.~\ref{fig:kin_emiles}. { The red line is obtained by performing a linear fit to all points within a galactocentric distance of $R=20$'' (i.e. $\rm \sim R_e/4$), and connecting this fit at $R \sim 20$'' to a constant value of $\sigma \sim 260$~\kms\ for larger galactocentric distances.}

\end{appendix}
\end{document}